\documentclass[twocolumn]{aastex63}
\begin{document}

\title{Photoionization Models for High Density Gas}

\correspondingauthor{T. Kallman}
\email{timothy.r.kallman@nasa.gov}

\author[0000-0002-5779-6906]{T. Kallman}
\affiliation{NASA Goddard Space Flight Center, Code 662, Greenbelt, MD 20771}

\author[0000-0002-5522-9705]{M. Bautista}
\affiliation{Department of Physics, Western Michigan University, Kalamazoo, MI}

\author[0000-0003-3828-2448]{J. Deprince}
\affiliation{Physique Atomique et Astrophysique, Universit\'e de Mons-UMONS, B-7000 Mons, Belgium}

\author[0000-0003-3828-2448]{J.~A.~Garc\'ia}
\affiliation{Cahill Center for Astronomy and Astrophysics, California Institute of Technology, Pasadena, CA 91125, USA}
\affiliation{Dr. Karl Remeis-Observatory and Erlangen Centre for Astroparticle Physics, Sternwartstr.~7, 96049 Bamberg, Germany}

\author[0000-0002-2854-4806]{C. Mendoza}
\affiliation{Department of Physics, Western Michigan University, Kalamazoo, MI}

\author[0000-0003-4504-2557]{A. Ogorzalek}
\affiliation{NASA Goddard Space Flight Center, Code 662, Greenbelt, MD 20771}
\affiliation{Department of Astronomy, University of Maryland, College Park MD}

\author[0000-0002-4372-6798]{P. Palmeri}
\affiliation{Physique Atomique et Astrophysique, Universit\'e de Mons-UMONS, B-7000 Mons, Belgium}

\author[0000-0002-3937-2640]{P. Quinet}
\affiliation{Physique Atomique et Astrophysique, Universit\'e de Mons-UMONS, B-7000 Mons, Belgium}

\begin{abstract}

Relativistically broadened and redshifted 
6.4 -- 6.9 keV iron K lines are observed from many accretion powered 
objects, including X-ray binaries and active galactic nuclei (AGN).
Existence of gas close to the central engine implies large radiation 
intensities and correspondingly large gas densities if the gas is to remain
partially ionized.
Simple estimates indicate that  high gas densities 
are needed to allow survival of iron against ionization.
These are high enough that rates for many atomic 
processes are affected by mechanisms related to interactions
with nearby ions and electrons.    
Radiation intensities are high enough that stimulated processes
can be important.  Most models currently in use for interpreting relativistic
lines  use atomic rate coefficients designed for use at low 
densities and neglect stimulated processes.  In our work so far we have presented 
atomic structure calculations with the goal of providing
physically appropriate models at densities consistent with line-emitting gas 
near compact objects.   In this paper we apply these 
rates to photoionization calculations, and produce ionization 
balance curves and X-ray emissivities and opacities 
which are appropriate for high densities and high radiation intensities.
The final step in our program will be presented in a subsequent paper:
Model atmosphere calculations which incorporate these rates
into synthetic spectra.

\end{abstract}

\section{Introduction}
\label{background}

Relativistically broadened and shifted iron K  
lines have now been observed from most of the 
galactic black hole X-ray binaries and active galactic nuclei (AGN) 
which are bright enough to allow detection, and from many
accreting neutron stars \citep{Cack10, Ludl20}.
This implies gas within 
$\sim$10 -- 100 gravitational radii ($R_G=GM/c^2=1.47 \times 10^5$ cm for $M= 1 M_\odot$)
of the compact object.  Variability timescales \citep{Uttl14} suggest that the 
source of continuum radiation comes from a region which is no larger 
than the region responsible for line emission.
Iron K line emission can be produced by reprocessing 
the strong X-ray continuum via inner shell fluorescence.
In this process, an X-ray photoionizes a K shell electron in an ion 
followed by a radiative line emission involving a transition of 
an L-shell electron to the K-hole.  Because of this mechanism, 
iron K line emission can be produced, 
with varying efficiency, by essentially any ion of iron with the 
exception of fully stripped, bare nuclei.  

The detection of the  K lines provides a limit on the degree of ionization 
in the gas responsible for the emission.  This can be described in terms of 
the ionization parameter,  $\xi=4\pi F/{\rm n}_e$, where $F$ is the local radiation 
flux and ${\rm n}_e$ is the gas density.  Photoionization models show that 
iron is fully ionized if $\xi \geq 10^5$ erg cm s$^{-1}$ for 
plausible choices for the spectral energy distribution (SED) \citep{Kall01}.
As an example, in the galactic black hole candidate LMC X-1 
approximately 90$\%$ of the line emission must come from within 10 $R_G$ of the compact object
\citep{Stein12}  and  the continuum luminosity and mass are $L\leq 1.8 \times 10^{38}$ erg s$^{-1}$, 
$M=$4 -- 10 $M_\odot$ \citep{Mccl06}.
Therefore the continuum flux in the line emitting region must be 
\begin{equation}\label{flux}
F\geq 10^{22}\ {\rm erg\,cm^{-2}\,s^{-1}}\ .
\end{equation}
\noindent This, together with the constraint on ionization parameter, implies that 
the density be ${\rm n}_e \geq 1.4 \times 10^{20}$ cm$^{-3}$.
This density estimate agrees well with the standard thin-disk model of
\cite{Shak73}. This model predicts that the density of a radiation-dominated
disk is ${\rm n}_e \propto m^{-1}_\mathrm{BH}\dot{m}^{-2}$, where $m_\mathrm{BH} =
M_\mathrm{BH}/M_{\odot}$, and $\dot{m}$ is the mass accretion rate in
units of the Eddington accretion rate $\dot{M}_\mathrm{Edd}$.  Thus, the disk
density at $R = 2R_g$ is ${\rm n}_e \sim
10^{16-22}$\,cm$^{-3}$, for maximally spinning black holes with masses in the
range $m_\mathrm{BH} = 10^6-10$, and $\dot{m} \sim 10$\% \citep[see also Figure
1 in ][]{Garc18}.

The detailed physical conditions in compact X-ray source accretion disks are affected by both electromagnetic and hydrodynamic processes \citep{Balb91}. Estimates for the gas density depend on detailed  multi-dimensional magneto-hydrodynamic (MHD) calculations and are not  simple functions of radius or height above the disk plane. As an example, MHD simulations  can produce densities $\geq$10$^{20}$ cm$^{-3}$ in the region close to the  disk plane \citep{Nobl10, Schn13} for a 10 M$_\odot$ black hole  accreting at 10$\%$ of the Eddington luminosity.

Models for fitting to observed relativistic lines begin with models  which provide the emitted spectrum from a localized region of the  accretion disk.  These  'reflection models' depend on local conditions: ionization parameter $\xi$,  elemental abundances, ionizing spectrum shape, electron number density and turbulence.  Rates for atomic processes affecting the ionization,  recombination, excitation and thermal balance are calculated and  used to solve the equations of local balance.   This determines ion fractions and level populations for all astrophysically abundant elements, as well as electron kinetic  temperature.  Line and continuum opacities and emissivities  are also calculated.  Typical calculations adopt plane parallel geometry  and one dimensional radiation transfer to calculate synthetic  X-ray spectra.   For the purposes of calculating the thermal balance and emitted spectrum of photoionized gases, accuracy and comprehensiveness of  rate coefficients for atomic processes are more important for  accurate modeling than more complex geometrical considerations or radiative  transfer algorithms \citep{Kwan81, Kall07}.

Widely used reflection models include those produced with the {\sc reflionx}
code by \citet{Ross05,Ross07}, and those produced with the {\sc xillver} code
by \citet{Garc10,Garc11,Garc13}. The latter uses the {\sc xstar} atomic
database and subroutines for the calculation of the ionization, excitation, and
thermal balance to determine the line and continuum opacities and emissivities.
{\sc
xstar}\footnote{http://heasarc.gsfc.nasa.gov/docs/software/xstar/xstar.html}
\citep{Kall01, Baut01} embodies a relatively complete and up-to-date treatment
of the atomic processes occurring in a photoionized gas, although it is
limited until now to densities ${\rm n}_e \leq 10^{18}$~cm$^{-3}$. 
The former employs atomic rate coefficients which are based 
on low densities approximations. 

The reflection model, and the line emission microphysics, 
are linked to determination of spin and other quantities.
Effects of spin are included in the model by summing the reflection spectrum 
over disk area and convolving with a transfer function which takes
into account relativistic effects \citep{Dauser14,Garcia14}.
Fits to observed lines appear to show that the best fit spin is correlated 
with the iron abundance \citep{Reyn12, Stein12}, and/or with the location of
the inner radius of the accretion disk \citep[e.g.,][]{Wang-Ji2018}.
Moreover, a large fraction
of relativistic lines in AGN require iron abundances 
greater than solar \citep{Reyn97,Garc18b}.  Iron abundance is typically the only 
free parameter available to account for uncertainty in the line 
reprocessing efficiency in most reflection models, and 
the inferred high abundances, and apparent correlation with spin, 
suggest that the current models 
underestimate the rate of iron line production.

The accuracy of the results from all reflection spectral fits are dependent on
systematic model limitations, in addition to any statistical uncertainties. The
unresolved question of the high iron abundance to fit the observational data
from accreting black holes, as described above, is perhaps the most obvious
indication of model systematic uncertainties.  
Recent theoretical works have explored the
effect of higher density grids when computing the X-ray reflection spectra of
accretion disks (still assuming a constant disk density) using the {\sc xillver}
code \citep{Garc16}. Calculations performed over a range of densities, have
demonstrated that at sufficiently high densities (${\rm n}_e > 10^{17}$~cm$^{-3}$),
ionization effects result in a significant increase of the atmospheric
temperature. However, limitations in the conventional atomic data prevented
the calculation of these models at densities above ${\rm n}_e = 10^{19}$~cm$^{-3}$.

The implementation of high density reflection models with current atomic data 
have already shown to have important consequences in the analysis
of AGN and black hole binaries (BHBs). In the case of the latter, these high density models have been
tested in two systems: Cyg~X-1 \citep{tom18} and GX~339$-$4 \citep{jia19}. In
both cases the iron abundance predicted by the improved model was significantly
decreased relative to results obtained with the standard, lower-density (${\rm n}_e =
10^{15}$\,cm$^{-3}$) disk reflection models.
In the case of AGN, the inclusion of high density effects could provide a new
physical explanation for the mysterious soft excess observed in the X-ray
spectrum of many Seyfert galaxies. Calculations presented in \cite{Garc16}
suggest that the soft excess is likely a measure of the density in the
accretion disk, which would transform it into a powerful diagnostic tool.
High density models have been successfully used to fit the X-ray spectra of the
AGN IRAS~13224$-$3809 \citep{jia18}, Mrk~1044 \citep{mal18}, and Mrk~509
\citep{gar19a}. A much larger study has been conducted by \cite{jia19b},
including a sample of 17 Seyfert 1 AGN with strong soft excess, using the same
new high density reflection models from \cite{Garc16}.
All these sources display strong soft excess components in
their spectra. Similarly to the results reported for BHBs, fitting the observed
soft excess in these AGN with high density reflection models resulted in a
lower (and more physical) iron abundance of the reflector \citep[see also
discussion in][]{par18}.

The physics of photoionization including densities beyond those typically 
assumed for AGN broad line regions has been discussed 
previously by  \citet{Rees89}.  Density effects in coronal models leading 
to the suppression of dielectronic recombination (DR)
have been addressed by \citet{Summ72}.  \citet{Dufr20} have 
explored the effects of DR suppression and metastable populations on 
ionization balance and line diagnostics in the solar transition region.

The list of processes which can be affected by plasma effects at 
the densities near compact objects includes:
(i) The effect of nearby ions on bound electrons which screens the nuclear 
charge and reduces the binding.  States 
with the largest principal quantum numbers can have their energy levels 
perturbed or can be unbound.  This 'continuum lowering' reduces the 
states which can  be counted when 
summing to calculate the net rate coefficients for recombination 
or any other process where high-$n$ states are involved.
(ii) Dielectronic recombination:  this process, whereby recombination occurs 
together with excitation of a bound electron in the target ion, is important 
particularly for photoionized plasmas.  During the recombination, the doubly 
excited ion is particularly fragile and subject to collisional ionization.  This
effectively suppresses the recombination rate at high density.
(iii)Three body recombination:  this process can be important for 
any bound level at high density.  It can dominate the net recombination rate 
and produces a different distribution of excited states 
than radiative recombination.
(iv) Stimulated radiative processes will become important when the radiation 
energy density is high at the energies of appropriate transitions, 
and will enhance the rates for radiative decay processes.
(v) Although metastable levels are already included in 
state-of-the-art calculations, at sufficiently high densities 
new levels can be collisionally populated from the ground, thereby 
affecting the emissivity and opacity.
(vi) At sufficiently high density, the effect of
neighboring ions and electrons can create plasma microfields that perturb
 the atomic level structure
in ways which can change which decay or excitation channels are energetically 
allowed, and can change atomic wavefunctions and the resulting matrix elements
and rate coefficients.  This is formally the same as item (i) but computationally 
it is treated differently owing to the way in which we treat high-$n$ states 
in contrast with lower excitation states. (vii) Free-free heating is strongest at high 
density and for illuminating spectra with flux at soft energies.  

In this paper we present models for photoionized gas which are appropriate for 
gas densities up to $\sim 10^{22}$ cm$^{-3}$. 
These are incorporated into the 
{\sc xstar} computer code and all results are derived from this code.  We have 
also incorporated new atomic data for odd-Z and low-abundance iron peak 
elements.   In \cite{Kall01} we presented 
a discussion which overlaps with some of what is presented here, along with 
models which could be applied to densities up to $10^{16}$ cm$^{-3}$.  This 
paper represents an extension of those models to higher density.  

A primary goal of this work is to explore the effect of high densities on the 
observable spectra from, for example, illuminated disks near black holes.  In this 
paper we lay the groundwork for such calculations by describing the physical 
processes and the ingredients which are incorporated into the {\sc xstar} code.
We include sample calculations of opacities and emissivities and simple 
single-zone spectra.  We defer the calculation of physical models for 
illuminated disks until a future publication.  In Section
\ref{basics} we describe the various effects of high density on atomic rates and
our approach to calculating them.  In Section \ref{results} we present results, 
and in Section \ref{discussion} we summarize.

\eject

\section{Photoionization Equilibrium Modeling at High Density}
\label{basics}

The microphysical processes in gas very close to a compact 
object are likely to be affected by the strong radiation 
field, and also by high gas density.
High plasma density  can affect many of the relevant atomic 
processes, either by truncating the bound 
levels with high principal quantum number (continuum 
lowering), increasing importance of collisional processes, or
changing the effective nuclear charge and hence
the atomic structure and associated rate coefficients.

It is worthwhile to discuss in more detail the context of the
models presented in this paper.  In our previous paper \citep{Kall01} we described 
the assumptions used to calculate photoionization models in gas up to density 
$\sim 10^{16}$ cm$^{-3}$.  Temperatures in the atomic or ionized gas found by these 
models range from $\sim 10^4$ to $\sim 10^8$K.  In this paper we extend those results 
to densities $10^{22}$ cm$^{-3}$.  These are limited by the assumption that the plasma
screening parameter $\mu=\sqrt{\frac{4 \pi {\rm n}_e^2}{kT}}$ be $\leq$0.5, which permits
the application of Debye theory to the treatment of the ionic level structure.  
These conditions can be compared with those found in terrestrial plasmas, such as 
laser-produced plasmas:  the conditions we treat are slightly below those described
as `High Energy Density' plasmas, which have P=1Mbar, and are also mostly above
the boundary of strongly coupled plasmas.  They are also slightly below the densities
of `Warm Dense Plasmas' \citep{weis06}.

\subsection{Atomic Data}

Although it is not the primary objective of this paper, this work 
coincides with a major updating of the atomic data for radiative 
and collisional rates for the odd-Z elements below $Z$=20 and the trace
elements above $Z$=20 in the {\sc xstar} database.  This work has been reported in 
\citet{mend18,mend17,palm16,palm12}.  
This results in a growth of the {\sc xstar} atomic
database by more than a factor 2.  Previously, the data for ions from 
isoelectronic sequences with 3 or more electrons relied on hydrogenic 
approximations for energy levels and rate coefficients.  The new data 
is the result of calculations using the HFR \citep{Cowa81}, Autostructure
\citep{Badn11} and GRASP \citep{Gran80} codes, intercomparing 
results from the various platforms to understand inconsistencies.
Photoionization cross sections have been calculated using the 
R-Matrix methods \citep{Berr95}.
The new data represents several $\times 10^4$ new lines and levels from 
all the ions with 3 or more electrons of 
F, Na, P, Cl, K, Ti, V, Mn, Cr, Co, Cu and Zn.

\subsection{Continuum Lowering Effect for High-$n$ States}
\label{continuumlowering}

Radiative recombination can occur onto levels with arbitrary principal quantum 
number $n$.  The slow $\sim 1/n$ decrease in the dependence of this rate on 
principal quantum number means the ensemble of high-$n$ states make a 
significant contribution to the total rate.
When the potential seen by a bound electron
becomes dominated by the surrounding ions rather than by the electron's own nucleus,
then the rates for decay to lower levels are likely to be slower than processes which 
connect the electron to the continuum.  If so, electrons should be treated as if 
not bound to the nucleus.  This continuum lowering provides an upper limit to the 
range of states with are considered bound to the ion.  Our treatment of this 
process and its effect on the recombination rates, and much of the discussion in this subsection, 
is the same as presented in \citet{Kall01} extended to higher densities.

 {\sc xstar} does not directly use 
the total recombination rate, but rather calculates rates onto a set of 
spectroscopic levels, typically with principal quantum numbers $n\leq$6, using photoionization 
cross sections and the Milne relation.  Then it calculates a photoionization cross section for
one or more fictitious superlevels.  The recombination rate onto the superlevel is calculated 
from the photoionization cross section via the Milne relation.  The total recombination rate 
onto the ion is the sum of the rate onto the superlevels plus the rates onto the spectroscopic 
levels.  The photoionization cross section
from the superlevel is chosen so that when it is used to calculate a recombination rate, and all 
the rates are summed, the total recombination rate for the 
superlevel(s) plus the spectroscopic levels adds to the total rate taken 
from one of the compilations by ADAS \citep{Badn03} where available and from \cite{Aldr73} otherwise.
 The R-Matrix photoionization cross sections for the spectroscopic levels include resonances, so their
application to recombination via the Milne relation therefore implicitly 
includes the inverse of the resonance ionization channel, which is 
dielectronic recombination. This does not result in double counting of these dielectronic 
recombination channels because the recombination rate onto the superlevels is chosen so that it, plus
the Milne rates onto the spectroscopic levels, sums to the correct total rate.
 The superlevels  decay directly to ground without the emission of any observable cascade radiation 
for ions with 3 or more electrons.  The
exception is the decay of the H- and He- isoelectronic sequences, for which we have explicitly calculated
the decay of the superlevels to the spectroscopic levels using a full cascade calculation
\citep{Baut98, Baut00}.   The superlevels are chosen to have energies close to the continuum.  
We include both radiative and collisional transitions to the ground level (and other levels in the case
of H- and He-like ions).  We include collisional coupling of the superlevels to 
the continuum for H- and He-like ions.  For other isoelectronic sequences
the superlevels decay only to ground and so will not lead to enhanced 
line emission. 
We apply density-dependent suppression factors to the recombination 
rates onto the superlevel in order to take into account density effects on radiative and dielectronic recombination.

Various detailed criteria can be used to describe the cutoff value of $n$ including:
Debye screening occurs when atomic binding 
is dominated by screening of the nuclear charge by nearby electrons.
Particle packing occurs when the mean internuclear separation 
in the plasma is smaller than the distance from the nucleus to the high-$n$
ionic orbitals.
Stark broadening occurs when the fluctuating microfield from 
nearby changes causes atomic levels to merge with 
each other \citep{Ingl39}.
The effective continuum level is set by the minimum of these criteria.

Estimates of the high-$n$ cutoff due to  particle packing can be made
based on when the mean inter-nuclear distance becomes smaller than 
the size of the atomic orbital.  If so, one can 
define the high-$n$ cutoff level as \citep{Hahn97}
\begin{equation}\label{np}
n_P=(1.9 \times 10^8 Z)^{1/2} {\rm n}_e^{-1/6} = 6.4 Z^{1/2} {\rm n}_{20}^{-1/6}
\end{equation}
where we use $n$ for the principal quantum number and ${\rm n}_e$ for the gas number density, 
n$_{20}$ is the number density in units of $10^{20}$ cm$^{-3}$, 
and $Z$ is the nuclear charge.

The importance of Debye screening  can be estimated 
using Debye-Huckel theory.  
The characteristic length is 
\begin{equation}
\lambda_D=\sqrt{\frac{kT}{4 \pi {\rm n}_e}}=2.4 \times 10^{-7} {\rm n}_{20}^{-1/2}T_4^{1/2} {\rm cm}
\end{equation}
where $T$ is the gas temperature and
$T_4$ is the gas temperature in units of 10$^4$ K.
This corresponds to  an atomic level 
near a nucleus with charge $Z$ with principal quantum number 
\begin{equation}\label{nd}
n_D=4.8 {\rm n}_{20}^{-1/4}T_4^{1/4} Z
\end{equation}

Under a high concentration of singly charged ions in a plasma, a microelectric 
field is formed that will lead to Stark broadening of the atomic levels. 
Then, for sufficiently high-$n$ numbers, the atomic levels will merge with 
each other, which lowers the continuum. In this case the continuum level
is given by \citep{Ingl39}
\begin{equation}\label{ns}
n_S=(1.8 \times 10^{26} Z^6/{\rm n}_e)^{2/15} = 6.8 Z^{4/5} {\rm n}_{20}^{-2/15}
\end{equation}
For temperatures lower than $10^5  (Z^2/{\rm n}_e)$K , the electrons contribute to the 
broadening through the static Stark effect. Therefore the density ${\rm n}_e$ in the 
Equation (\ref{ns}) should include both positive and negative charges. At higher 
temperatures the electrons contribute to the broadening by means of collisions, 
but this is smaller than the Stark effect of the same electrons at lower 
temperatures. At high temperatures only positive charges are considered for 
this equation.

The effective continuum level $n_C$ is the minimum of $n_P$, 
 $n_D$, and $n_S$.  
In the case of oxygen, for example, under conditions of $T = 10^5$ K and 
${\rm n}_e = 10^{16}$ cm$^{-3}$, 
only levels with $n\leq $83 are bound, while at ${\rm n}_e = 10^{20}$ cm$^{-3}$ only 
levels with $n \leq$ 18 are bound 
\citep{Baut98}.
The scaling of the various cutoffs with plasma conditions are apparent 
from equations \ref{np}, \ref{nd}, \ref{ns}:  at high density particle packing provides 
the lowest limit to the allowed principal quantum number; at low density and 
low Z Stark broadening dominates; Debye screening dominates at low temperature, low Z and high 
density.  We note that none of these expressions is intended to be applied at densities
beyond 10$^{22}$ cm$^{-3}$, where ion-sphere or equivalent expressions are more 
appropriate.
We have implemented the above expressions leading to $n_C$, and then 
used them for a summation of the hydrogenic rate coefficients.  These 
are used to calculate a fraction,  which we call a suppression factor, which is 
the sum of the rates over $n$ up to $n_C$ divided by 
the sum of the rates over $n$ up to a typical low density cutoff of 200.  This 
suppression factor is 
then applied to standard low density radiative recombination rate coefficients 
from ADAS \citep{Badn03} where available and from \cite{Aldr73} otherwise.

All of the scaling relations contributing to the estimate for $n_C$ 
are based on hydrogenic energy levels.  Therefore estimates of the expressions for the cutoff
$n$ are only valid for high-$n$ states.   Furthermore, we only apply the suppression factor 
to the superlevel, so this treatment alone can never reduce the radiative recombination rate below the 
summed rates onto the spectroscopic levels.  In Section \ref{debyehuckel} we describe 
how we treat density effects at densities such that the $n$ cutoff affects the spectroscopic levels.

\subsection{Dielectronic Recombination}

Dielectronic recombination (DR) occurs when a recombination 
event is accompanied by an excitation of the recombining 
ion. The resulting doubly excited ion can be re-ionized 
by collisions instead of decaying.   These effects 
have been discussed and modeled beginning 
with \cite{Summ72}.  
\cite{Badn93} showed that low density  dielectronic recombination rates 
in widespread use produce ionization balance curves for oxygen
which are inaccurate by large factors at densities ${\rm n}_e \sim 10^{13}$ cm$^{-3}$.
Dielectronic recombination separates into high temperature
and low temperature.
In the former case, the core excitation
involves a change in principal quantum number $n$, and 
the captured electron is in a high-$n$ state, while 
in the latter case the core excitation has $\Delta n$=0 
and the captured electron is in a state with lower $n$
(hydrogenic ions and those with closed shells cannot 
participate in this process).
The high temperature case is expected to be more
susceptible to density effects.
\cite{Niko13} provide convenient expressions for the effect of
finite density on dielectronic recombination, applicable 
to  DR at densities up to ${\rm n}_e = 10^{20}$ cm$^{-3}$.
These expressions are based on extrapolating the results 
of \citet{Badn93} across densities, and also applying 
simple rules for which doubly excited levels are likely to be
produced for various isoelectronic sequences.  
This allows them to  supply scaling 
expressions for the  suppression of DR due to electron collisions.
They predict that dielectronic 
recombination will be almost entirely suppressed for many
ions at the highest densities.
We incorporate these into our treatment of this 
process for all ions, as it affects the rates into the superlevels 
described in the previous subsection.  At density ${\rm n}_e \sim 10^{20}$ cm$^{-3}$ 
DR suppression renders it negligible and we use the value at that density for all higher values.

\subsection{Three Body Recombination}

According to a semi-classical formulation, three body recombination occurs when an 
electron approaches an ion with kinetic energy greater than the 
binding energy of the recombined level
and  that a second electron be within the same volume in order to 
carry away the liberated energy.  
Three body recombination is the inverse of collisional ionization.
Three body recombination 
coefficients increase with the principal quantum number 
roughly as $n^3$.  Radiative recombination coefficients 
scale with the principal quantum number roughly as $n^{-1}$. Therefore, the 
contribution of three body recombination to the total recombination process 
will shift toward higher excitation states as the nuclear charge of the ion 
increases. 
For example, at temperature $T=10^5$ K and density ${\rm n}_e = 10^{20}$ cm$^{-3}$, 
three body rates dominate over radiative rates for direct recombination, 
not considering cascades, of levels with  $n\geq$2 for oxygen, and $n\geq$3 
for neon.  
Three body recombination has a rate per unit volume scales with density 
$\propto {\rm n}_e^3$, while radiative and dielectronic recombination scale 
$\propto {\rm n}_e^2$.  Thus three body gains in importance at high density.
The importance of three body recombination rates 
increase with respect to the radiative rates as temperature decreases.
The total recombination rate onto the H-like 
ion and onto the $n$=2 states by means of the cascades from higher states 
will be dominated by three body recombination at temperatures 
below approximately 3 $\times 10^5$ K for density ${\rm n}_e = 10^{17}$ cm$^{-3}$. 
Three body recombination will affect the 
ionization balance of the plasma and the emitted recombination spectrum. 
As three body recombination occurs preferentially onto highly excited levels, 
the line emission from these levels will be strongly enhanced.
In hydrogenic ions, Lyman-series line emission 
is also enhanced due to the cascades from high levels onto the $n=2$
state multiplet.

Three body recombination can affect all levels at high densities.  
We implement this process for all levels.
This requires collisional ionization cross sections; we use 
values from the literature for ground states and excited states where available
\citep{Baut01}.
We use hydrogenic rates for excited levels where other rates are not available. 
The excited levels which are treated explicitly 
in the calculations by ourselves and others are typically 
those which can be most easily excited from the ground level by 
electron collisions, supplemented by those which can 
be populated by radiative recombination.  
In our case, these come primarily from 
the {\sc chianti} collection \citep{Land13}.
Three body recombination can populate levels with a different 
set of properties:  large collisional coupling to the continuum.

Our total recombination rates also implement the Nikolic \citep{Niko13} DR suppression multiplying
the ADAS recombination rates \citep{Badn03}, new cascade
 calculations for H- and He-like ions which are valid up to densities
${\rm n}_e = 10^{22}$ cm$^{-3}$ and the continuum lowering treatment which was used since
Paper I.  We also  include three body recombination using the inverse of our
 collisional ionization rates, which are mostly hydrogenic,  for the
 ground and most singly excited levels.

\subsection{Stimulated Processes}

Stimulated processes become important when the  radiation intensity approaches the value for  a blackbody at the local temperature.  This  ratio can be expressed as the photon occupation  number $F_\varepsilon (2\varepsilon^3/h^3c^2)^{-1}$ where $F_\varepsilon$ is the monochromatic intensity at  energy $\varepsilon$.  This is approximately  $F \varepsilon_{keV}^{-4} 1.5 \times 10^{-23}$ where $F$ is  the total intensity (in erg cm$^{-2}$ s$^{-1}$)  and $\varepsilon_{keV}$ is the photon  energy in keV.  Using the estimated flux from Equation \ref{flux},  shows that this quantity can be greater than unity.   This is most likely to be true for  $\varepsilon_{keV}\leq 1$, corresponding to valence shell recombination. An accurate assessment of the importance of stimulated  processes can only come from a detailed calculation, owing  to the sensitivity of the occupation number to photon energy.

Most calculations of line formation treat line  transfer using an escape probability formalism; this  implicitly takes into account stimulated line emission by  thermal radiation generated within the gas.  It does  not take into account stimulated emission from external  photons.  Stimulated recombination can enhance recombination  rates, which can affect inferences about the survival of iron ions against photoionization in accretion flows. Stimulated bound-bound decay can enhance line emission. Stimulated Compton scattering can affect the Comptonized spectrum but not the electron heating or cooling \citep{Sazo01}.

Stimulated processes are straightforward to  include in a calculation of level populations;  rates for recombination and radiative decay  are enhanced by a factor  $1+F_\varepsilon/\frac{2\varepsilon^3}{h^3c^2}$ at each energy in the integral expression for  the net rate.  This has been incorporated in the {\sc xstar} code, but only for recombination. The effects on bound-bound radiative decay employ  an escape probability treatment.

\subsection{Atomic Structure at High Density}
\label{debyehuckel}

The scaling relations for the high-$n$ cutoff of the total radiative recombination rates described in Section \ref{continuumlowering} are based on the assumption that the high-$n$  levels involved are hydrogenic.  These corrections are applied only to the  recombination rate onto the superlevel, i.e. the fictitious level which represents the highly excited levels which are not explicitly treated in our multilevel  calculation.  The recombination rates into levels other than the superlevel,  i.e. the spectroscopic levels,  are unaffected by the recombination rates described in  Section  \ref{continuumlowering}.

We take into account density effects on the spectroscopic levels by implementing the  results of  atomic structure calculated  by \citet{Depr20a,Depr19a,Depr19b}.  These authors carried out ab initio calculations of the structure of all stages of oxygen and iron ions using the  Multi-Configuration Dirac-Fock code together using a time-averaged Debye-H\"uckel potential to represent the plasma effects.  They  calculated the atomic structure  i.e. the energy levels, the first ionization  potentials (IPs), the K-thresholds, the wavelengths, and the decay radiative and Auger rates. They showed that the largest effect of the Debye-H\"uckel potential at high density  is on the lowering of the IP.  Lowering is important not only for the IP but also for the  K-threshold and for all other thresholds.  This will have  consequences for ionization balance and  also for opacities. The energy level structure and rates affecting inner shells are not  strongly affected by densities in the range that we consider here. In Debye theory the effects of screening are parameterized by the  plasma screening parameter 
\begin{equation}
\mu=\sqrt{\frac{4 \pi {\rm n}_e^2}{kT}}\ ,
\end{equation}
where $T$ is the electron kinetic temperature and ${\rm n}_e$  is the electron density.  The Debye-H\"uckel theory is valid for a  weakly coupled plasma, i.e. for a plasma coupling parameter  $\Gamma < 1$ and defined as 
\begin{equation}
\Gamma = \frac{e^2}{(4 \pi \epsilon_0 d k T)}
\end{equation} 
with $d = (3/4\pi {\rm n}_e)^{1/3}$ for a completely ionized hydrogen  plasma (with plasma ionization $Z^*=1$).
In a plasma  with gas number density ${\rm n}_e =10^{20}$ cm$^{-3}$ and temperature $T=10^4$K, then $\mu\simeq 0.1$.  Thus the results in this paper are applicable for densities up to ${\rm n}_e\sim 10^{22}$ cm$^{-3}$ for $T \sim 10^6$K.  The examples we give in the next section focus on these conditions. 

The calculations show that the change in the first ionization potential obeys a relatively simple scaling relation which is \citep{Depr20b}:
\begin{equation}\label{ip}
\Delta{\rm IP (eV)}\simeq -26.30  \mu Z_{eff} 
\end{equation}
where $Z_{eff}=Z-N+1$ and $Z$ is the nuclear charge and $N$ is the number of  bound electrons.  
The K-threshold and other thresholds obey a similar 'universal'  formula, i.e. 
\begin{equation}
\Delta E_K (eV) = -27.28 ~ \mu Z_{eff}
\end{equation}
These relations are shown graphically in \cite{Depr20b}.
It is important to  note that the absolute value of the IP lowering increases $\propto Z_{eff}$ while the ionization parameter itself scales crudely $\propto Z_{eff}^2$. Therefore the relative importance of IP lowering, measured as  $\Delta({\rm IP})/{\rm IP}$, is greatest for small $Z_{eff}$, i.e. low $Z$ and nearly neutral ions.

We implement the ionization potential lowering calculated by  \citet{Depr20a,Depr19a,Depr19b} in {\sc xstar} by changing the ionization potentials for all bound levels according to Equation (\ref{ip}).  The excitation energies relative to ground are not changed.
However, levels for which the excitation energy is greater than the 
new lowered ionization potential are excluded from the calculation, 
as are all processes connected to those levels.  The most important 
effect of this is to suppress recombination, since most recombinations
are to highly excited levels.   
{\sc xstar} calculates recombination using the 
Milne relation for most bound levels, plus additional recombination 
into a fictitious 'superlevel' which is close to the continuum.
IP lowering 
often results in truncation of the superlevel, with the accompanying 
decrease in the total recombination rate.
This procedure has the drawback that it is done in preprocessing, 
so we must assume  a single value of the screening parameter $\mu$ for an entire model.  


\subsection{Free-free Heating}

We have also implemented free-free heating in {\sc xstar}.  This process may lead to
strong heating at high densities, but the amount depends on the
shape of the incident radiation spectrum.  Both free-free and Compton heating depend on averages over the illuminating SED.  It is easy to show that the 
temperature (in units of $10^4$K) below which free-free heating exceeds Compton heating is:
\begin{equation}
T_4=4.04 \times 10^{-14} Z^4 g^2 {\rm n}_e^2 \left( \frac {<\varepsilon^{-3}>} {<\varepsilon>-4kT}  \right)^2 
\end{equation}
where $Z$ is the ion charge, $g\sim1$ is the Gaunt factor, ${\rm n}_e$ is 
the gas number density, and we have assumed a power law spectrum of 
radiation responsible for both the free-free and Compton heating.  
$\alpha$ is the slope of the ionizing spectrum 
in energy units, $\varepsilon_{min, eV}$ is the minimum energy of the ionizing
power law, and $<\varepsilon>$ is the mean photon energy.  Inserting 
plausible values for these parameters:  
shows that 
free-free heating can affect the gas up to the Compton temperature, $T_{IC} \geq 10^7$K
for densities ${\rm n}_e\geq 10^{10}$ cm$^{-3}$.  In what follows we adopt a spectrum
which has a lower level of free-free heating; we take a power law which is 
effectively cut off below 13.6 eV.  With this  spectrum the temperature 
at high ionization parameter, where only free-free and Compton processes are 
important, is $T = 9.1 \times 10^7$K at a gas density ${\rm n}_e=10^{20}$ cm$^{-3}$.  
This can be compared with the Compton temperature for this spectrum 
which is $T_{IC}=5.6 \times 10^7$K.  This choice of spectrum is motivated by a desire
to illustrate the effects of the other processes which affect gas at high 
density besides free-free heating; a spectrum with a single power law extending
to $\sim$0.1 eV can have an equilibrium temperature $T \geq 10^9$K  under 
comparable conditions.  Furthermore, in many AGN the observed 
flux at energies below $\sim$a few eV is likely reprocessed at large 
distances from the center, and the true soft flux incident on gas close 
to the center is very uncertain (eg. \citet{Deve13}).  Nevertheless, 
under a range of plausible conditions, free-free heating can have a 
significant quantitative 
effect on photoionization models for AGN \citep{Ogor20}.
The importance of this process in photoionization models at densities 
up to ${\rm n}_e \sim 10^{13}$ cm$^{-3}$
has also been pointed out by \citet{Rees89}.

\subsection{Photoionization Models}

The work presented in this paper  incorporates the corrections to atomic 
rate coefficients described so far into calculations 
of line reprocessing using the {\sc xstar} package.
It calculates ionization, temperature, opacity and 
emissivity of X-ray gas self-consistently.  
It is in widespread use for synthesizing X-ray, UV and 
optical spectra of astrophysical sources where photoionization is important
and for analysis of high resolution X-ray data.  
The code and atomic database, along with the {\tt warmabs/photemis} analytic models for {\sc xspec}, are freely available as part of the HEASoft\footnote{\url{https://heasarc.gsfc.nasa.gov/lheasoft/}} {\tt ftools} package. The standard {\sc xstar} distribution\footnote{\url{https://heasarc.gsfc.nasa.gov/xstar/xstar.html}} includes all elements with $Z \leq 30$.

The atomic data used by {\sc xstar} is taken from published sources,
including our own, and is freely available via our website.
These data  include: energy levels, line wavelengths, oscillator
strengths, photoionization cross sections, recombination rate coefficients,
and electron impact excitation and ionization rates \citep{Kall07}.
Atomic data for Li, Be, and B makes use of hydrogenic 
scalings for most quantities.

Results include calculations 
of general purpose ionization balance and temperature curves, 
 for various densities 
up to those  appropriate for relativistic lines near compact objects.
These can be used to calculate the emissivity and opacity for X-ray lines and continuum
as a function of ionization parameter and density. 
The implementation of the high density atomic rate coefficients will 
be included in the standard public {\sc xstar} distribution.

\section{Results}
\label{results}

\subsection{Recombination}

\begin{figure*}[p] 
\includegraphics*[angle=270, scale=0.6]{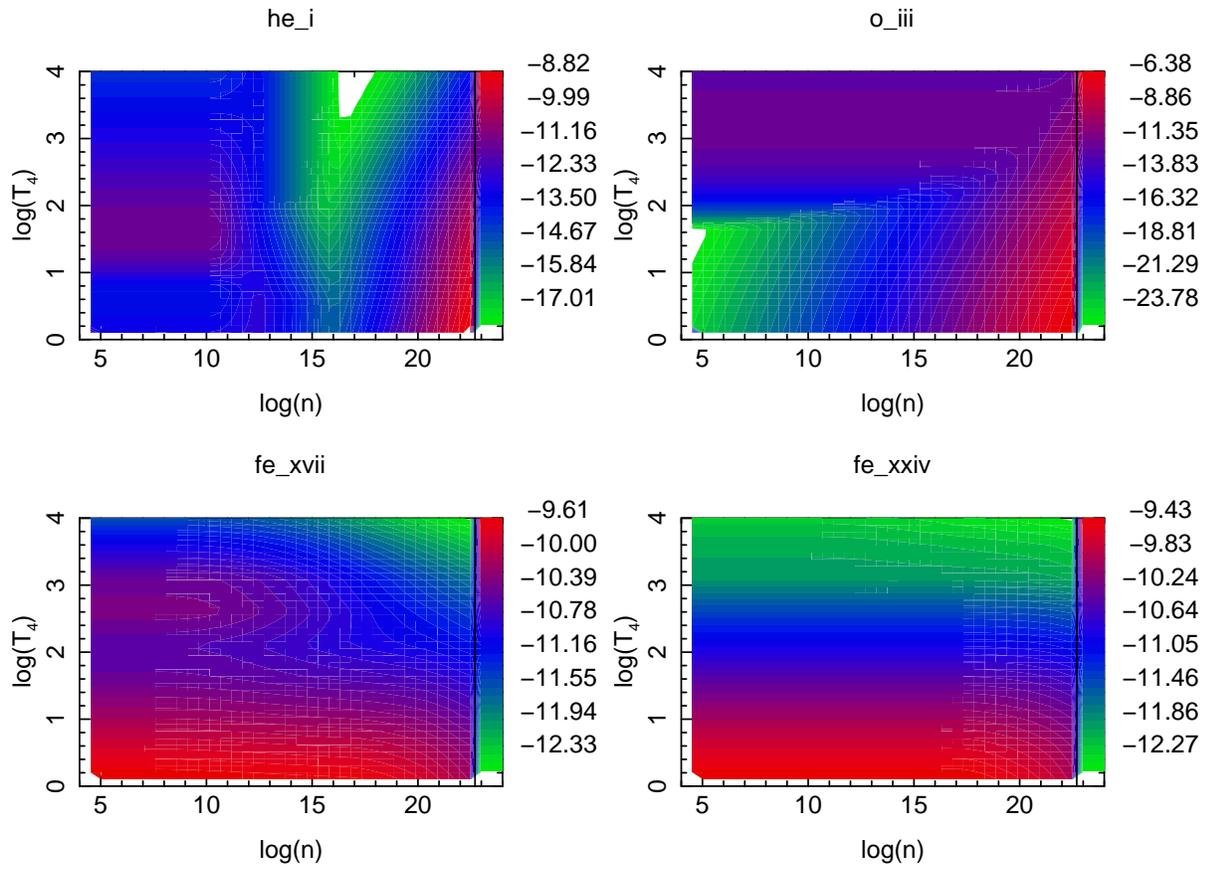}
\caption{\label{fig1} Sample recombination rates vs temperature and density.  $T_4$ is the temperature in units of 10$^4$K.  Color represents total rate coefficient in cm$^3$ s$^{-1}$}
\end{figure*}

Figure \ref{fig1} shows the total recombination rates used by {\sc xstar} as functions of temperature and 
density for sample ions.  Rates (units cm$^3$ s$^{-1}$) are shown as colors with logarithmic values shown 
in the color bars on the right side.  Many of the features of these plots are familiar:  at low density,
the rates are globally decreasing with temperature, which is due to radiative recombination, and 
with  a local maximum near ${\rm log}(T_4) \sim 1 - 2$ due to dielectronic recombination.  
At higher densities, it is apparent that the dielectronic recombination bump becomes weaker
and disappears between density ${\rm n}_e \sim 10^{10}$ -- $10^{15}$ cm$^{-3}$.  At still higher 
densities the effect of three body recombination is apparent in the strong 
increase in recombination rate above density ${\rm n}_e \sim 10^{18}$ cm$^{-3}$.
The different ions display the dependence on atomic number and ionization stage:
For He II the DR bump occurs at ${\rm log}(T_4) \simeq$ 1.5 -- 2 at low densities; the maximum 
suppression of recombination occurs near density ${\rm n}_e \sim 10^{15}$ cm$^{-3}$; three body 
recombination becomes important above density ${\rm n}_e \sim 10^{18}$ cm$^{-3}$.  For Fe XVII the low density 
behavior is dominated by the DR bump.  This is gone at densities ${\rm n}_e \geq 10^{15}$ cm$^{-3}$.  
For Fe the three body recombination comes in at higher densities than for lower-Z elements.
For Fe XXIV there is little density dependence in the total recombination rate.

\subsection{Ionization Balance}

\begin{figure*}[p] 
\includegraphics*[angle=0, scale=0.8]{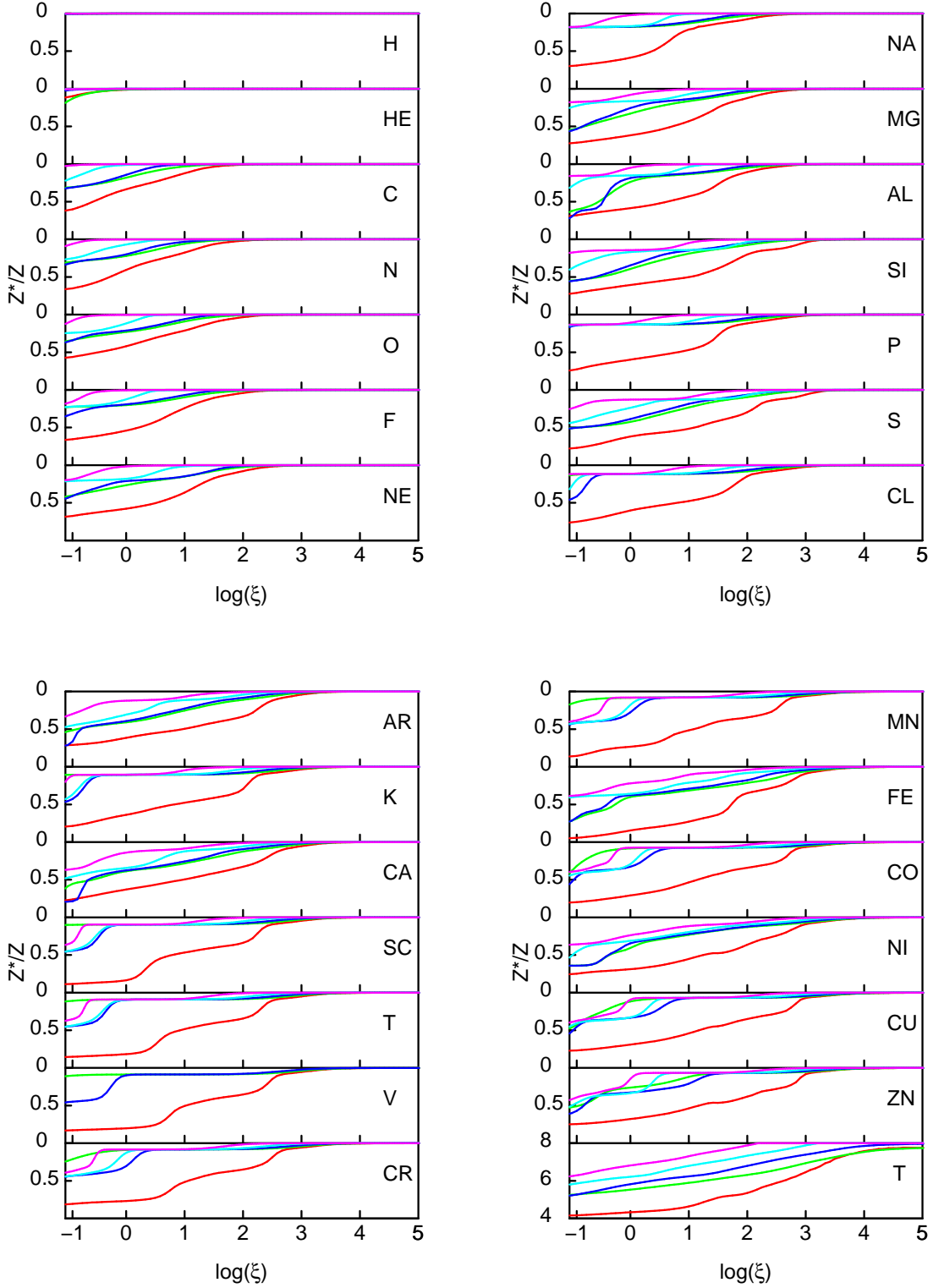}
\caption{\label{fig2} Mean ion charge and log(temperature) vs 
ionization parameter $\xi=L/nR^2$.  Colored curves correspond to  
density: 10$^4$  cm$^{-3}$ (red),
$10^{19}$  cm$^{-3}$ (green), $10^{20}$  cm$^{-3}$ (blue),  
$10^{21}$  cm$^{-3}$ (turquoise) and $10^{22}$ cm$^{-3}$ (purple).
Mean charge is displayed as fractions of the element atomic number.}
\end{figure*}

Figure \ref{fig2} shows the ionization balance for an optically thin 
gas vs. ionization parameter at various densities shown as colors ${\rm n}_e =10^{4}$, $10^{19}$, $10^{20}$, $10^{21}$, $10^{22}$ cm$^{-3}$.  
The low density ${\rm n}_e=10^4$ cm$^{-3}$ results display
the general properties of these models: (i) lower ion stages 
predominate at lower ionization parameter;  (ii) higher Z elements are 
more resistant to ionization and so their mean charge state is lower 
for a given ionization parameter than for low-Z elements; 
(iii) the temperature ranges from the Compton temperature $\geq 10^7$K at 
high ionization parameter 
to $\sim 10^4$ K at low ionization parameter where the gas becomes neutral.

At higher densities ionization potential lowering results 
in the truncation of the levels down to an energy which is a significant 
fraction of the low density ionization potential in this case.  For example,
for H at a density 10$^{20}$ cm$^{-3}$, the 
ionization potential is lowered by 1.8 eV.  This results in lower net 
recombination into all ions.    Figure \ref{fig2} 
shows that this has a significant effect throughout parameter space, and
that essentially all elements are more highly ionized than when this 
effect is neglected under the same conditions, owing to the reduced 
net recombination rate.  The temperature is higher 
owing to free-free heating and reduced collisional cooling.

\begin{figure*}[p] 
\includegraphics*[angle=0, scale=0.6]{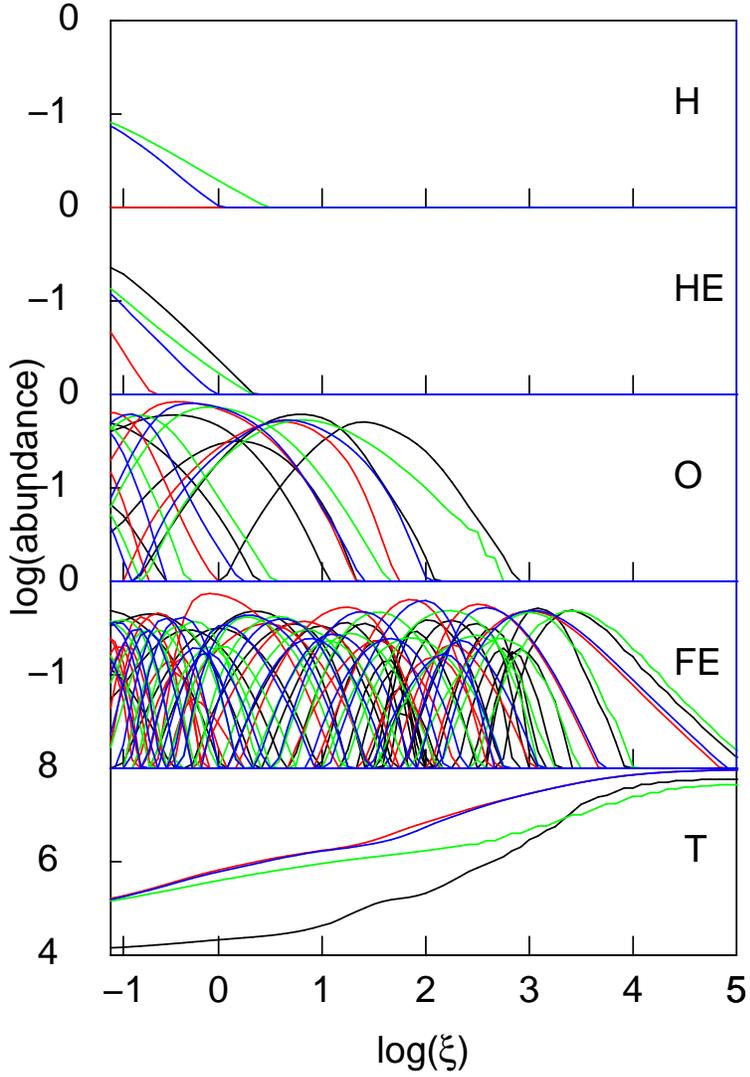}
\caption{\label{fig3} Equilibrium Ion fractions and temperature vs  ionization parameter  for selected elements, H, He, O, Fe, and temperature.  Black curve: density ${\rm n}_e = 10^{4}$ cm$^{-3}$,  other curves:  density ${\rm n}_e = 10^{20}$ cm$^{-3}$.  Red curve: including all density effects including ionization potential lowering on atomic structure; blue curve:  same as red but not including ionization potential lowering on atomic structure; green curve:  same as blue but not including the effects of free-free heating}
\end{figure*}

More detail about the differences between low and high density are shown in Figure
\ref{fig3}.  This compares the ion fractions for selected elements and temperature for 
models with low and high density and for high density models with and without some of the 
high density ingredients.  The red curve corresponds to high density with all the ingredients
in Section \ref{basics} included; the blue curve omits the change in ionization potential; 
the green curve omits free-free heating.  The black curve is the low density curve.  This shows
that the effect on temperature is to increase the temperature at high density by factors $\geq$ 10 
at low ionization parameter.  Free-free heating is important at high ionization parameter, and 
increases the temperature by factors $\geq$2.  This result depends on the shape of the 
illuminating SED, particularly at low energies, and here we have chosen an SED which is weak 
at low energies.  It is clear that most elements are more highly ionized at high densities, 
due to both the suppression of recombination and also to collisional ionization and 
reduced recombination due to higher temperatures.  The effect of IP lowering, i.e. the difference
between the red and blue curves is most apparent for the low-Z elements; IP lowering
results in much higher ionization of these elements owing to the fact that the level list
for most ions is truncated and recombination is greatly reduced due to the omitted levels.
The dependence of this effect on the effective ion charge is apparent in the iron 
ion fraction distribution:  the low charge states show a much greater difference between the 
red and blue curves while the highest charge states show very little difference.

\subsection{Heating-Cooling}

\begin{figure*}[p] 
\includegraphics*[angle=270, scale=0.6]{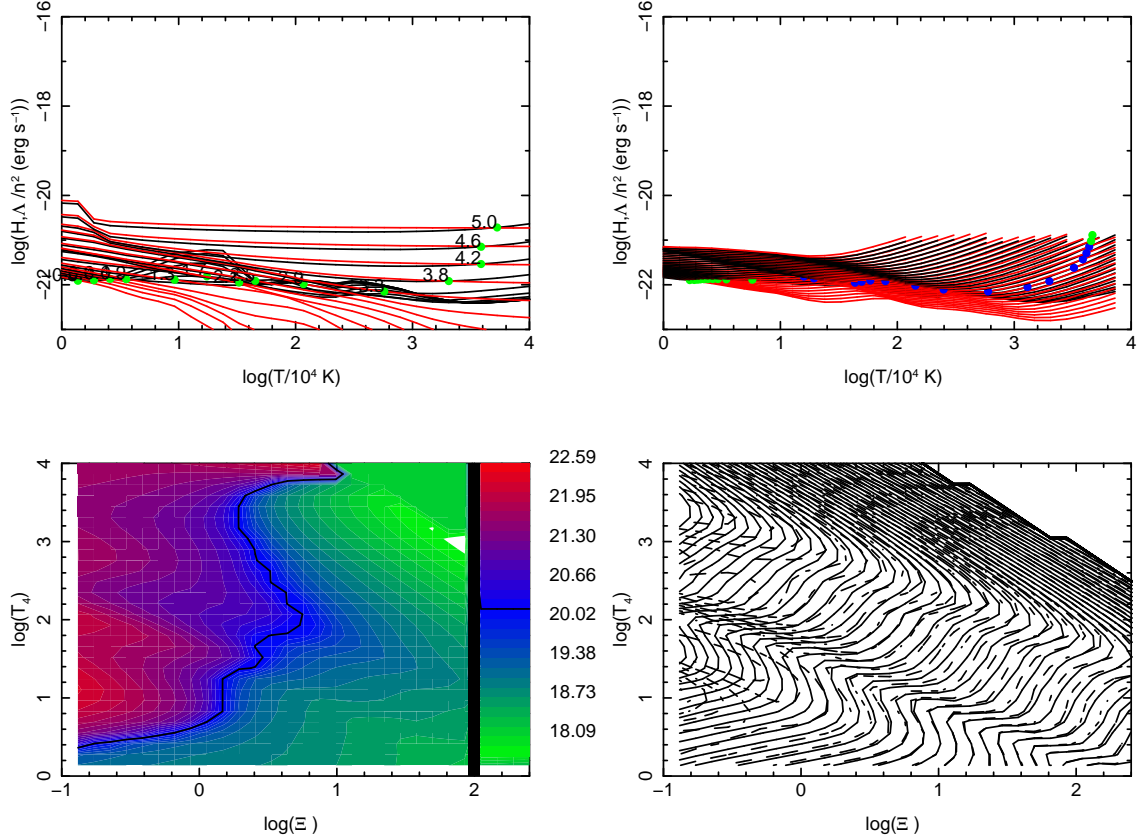}
\caption{\label{fig4} Heating and cooling rates vs. ionization parameter and temperature at density ${\rm n}_e =10^{4}$ cm$^{-3}$. Upper two panels show 
heating and cooling.  Different curves correspond to different ionization parameter, 
red=heating, blue=cooling.  Green dots correspond to equilibrium values.
Upper left panel is  for gas which is held at constant density, upper right 
panel is for gas held at constant pressure.  The bottom panels show 
heating and cooling values in the ($\Xi$-T) plane.  On the lower left the 
values are shown as colors, where the color value 
corresponds to ${\rm log}(H/{\rm n}_e^2-\Lambda/{\rm n}_e^2)+20$ for 
$(H/{\rm n}_e^2-\Lambda/{\rm n}_e^2)\geq 0$ and $-{\rm log}(\Lambda/{\rm n}_e^2-H/{\rm n}_e^2)+20$ for 
$(H/{\rm n}_e^2-\Lambda/{\rm n}_e^2)\leq 0$.  The black curve shows the equilibrium temperature, 
which has a characteristic `S-shape' indicative of thermal instability.
The lower right panel shows contours of constant heating and cooling separately in the 
($\Xi$-T) plane.  Dashed curves are heating, solid are cooling.}
\end{figure*}

Heating and cooling, and therefore the equilibrium temperature,  are affected 
by density.  At low density, electron impact collisions remove energy 
from the electron thermal bath and thus produce cooling.  The rates per 
unit volume scale with  density ${\rm n}_e$ as ${\rm n}_e^2$.  Electron impact 
cooling produces a characteristic local maximum in the cooling function 
vs. temperature, at temperatures $\sim 10^5$ -- $10^6$K, due to the presence 
of atomic transitions with energies $\sim 100$ -- 1000 eV.  This bump is 
stronger when the gas is at lower ionization, i.e. when photoionization 
is weak.  Heating is due to  photoionization, which produces fast 
photoelectrons which heat as they slow down by scattering with 
thermal electrons, and also Comptonization which heats
due to recoil.  Heating rates per unit volume therefore scale as $F{\rm n}_e$ where 
$F$ is the net flux.  When both heating and cooling rates are divided by density ${\rm n}_e^2$, to give the rates per particle, the heating rate is proportional
to the ionization parameter $\xi=4 \pi F/{\rm n}_e$.  Figure \ref{fig4} displays this
behavior.  The upper left panel shows the net heating and cooling rates
$H, \Lambda$ per particle vs temperature for various ionization parameters
for the photoionization conditions corresponding to the lowest density shown in Figure \ref{fig2}, 
i.e. density ${\rm n}_e = 10^4$ cm$^{-3}$.  
Cooling is black, heating is red. Green dots show the equilibrium 
temperatures.  The bump in the cooling curve above 
$10^5$K is apparent, and also the fact that the bump weakens at high $\xi$.
The cooling curves have a net increase with temperature, with the exception
of the region near the bump, owing to processes which are smoothly varying 
with temperature including bremsstrahlung and radiative recombination.
Higher temperatures result in reduced recombination and 
increased collisional ionization and therefore net higher ionization 
of the gas.  This results in reduced photoelectric  heating at higher 
temperature, and this is apparent in the red curves.  At higher ionization 
parameter Compton heating and cooling dominate and the curves converge 
over much of the parameter space.  At temperatures $\geq 10^8$K, 
which is not apparent in this Figure,  the Compton cooling increases 
$\propto T$.  Other panels in Figure \ref{fig4} show the behavior of 
heating and cooling at low density for gas which is held at constant pressure.
If so, the relevant ionization parameter is $\Xi=F/(cP)$ \citep{Krol81} where 
$P$ is the gas pressure.  The upper right panel shows the heating and cooling
per particle vs. T for various values of $\Xi$.  The bottom panels show 
heating and cooling values in the ($\Xi$-T) plane.  On the lower left the 
values are shown as colors, where the color value 
corresponds to ${\rm log}(H/{\rm n}_e^2-\Lambda/{\rm n}_e^2)+20$ for 
$(H/{\rm n}_e^2-\Lambda/{\rm n}_e^2)\geq 0$ and $-{\rm log}(\Lambda/{\rm n}_e^2-H/{\rm n}_e^2)+20$ for 
$(H/{\rm n}_e^2-\Lambda/{\rm n}_e^2)\leq 0$.  The black curve shows the equilibrium temperature, 
which has a characteristic `S-shape' indicative of thermal instability.
The lower right panel shows contours of constant heating and cooling separately in the 
($\Xi$-T) plane.  Dashed curves depict heating, solid  cooling.

\begin{figure*}[p] 
\includegraphics*[angle=270, scale=0.6]{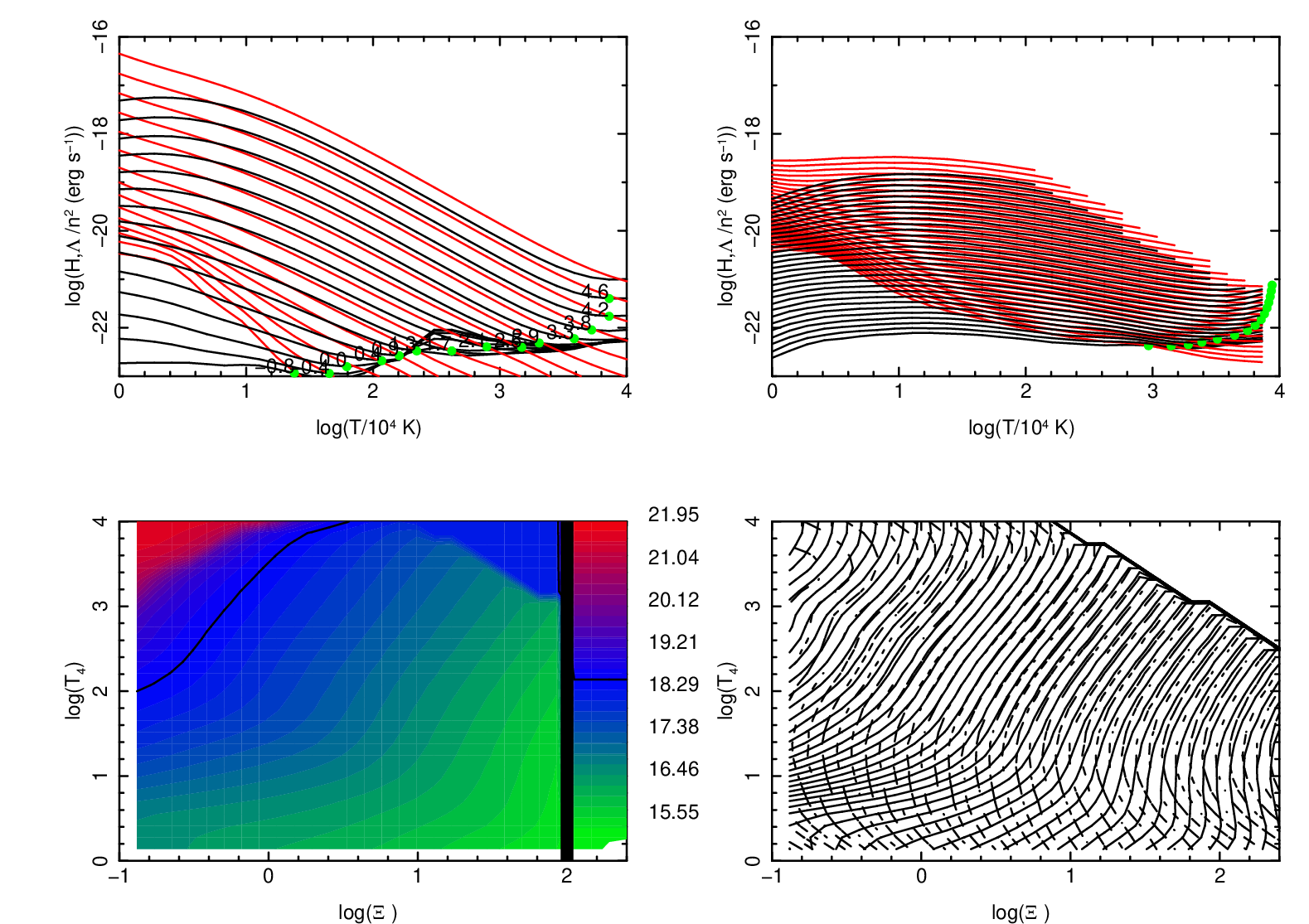}
\caption{\label{fig5} Heating and cooling rates vs. ionization parameter and temperature at density ${\rm n}_e =10^{20}$ cm$^{-3}$.  Upper two panels show 
heating and cooling.  Different curves correspond to different ionization parameter, 
red=heating, blue=cooling.  Green dots correspond to equilibrium values.
Upper left panel is  for gas which is held at constant density, upper right 
panel is for gas held at constant pressure.  The bottom panels show 
heating and cooling values in the ($\Xi$-T) plane.  On the lower left the 
values are shown as colors, where the color value 
corresponds to ${\rm log}(H/{\rm n}_e^2-\Lambda/{\rm n}_e^2)+20$ for 
$(H/{\rm n}_e^2-\Lambda/{\rm n}_e^2)\geq 0$ and $-{\rm log}(\Lambda/{\rm n}_e^2-H/{\rm n}_e^2)+20$ for 
$(H/{\rm n}_e^2-\Lambda/{\rm n}_e^2)\leq 0$.  The black curve shows the equilibrium temperature, 
which has a characteristic `S-shape' indicative of thermal instability.
The lower right panel shows contours of constant heating and cooling separately in the 
($\Xi$-T) plane.  Dashed curves are heating, solid are cooling.}
\end{figure*}

As indicated in Figure \ref{fig2}, at the highest densities three body 
recombination dominates over other recombination processes.  Three body 
recombination produces net heating since the third electron carries away 
the energy liberated in the process.  This is apparent in 
Figure \ref{fig5}, which shows heating and cooling at 
density  ${\rm n}_e =10^{20}$ cm$^{-3}$.  Comparison with Figure \ref{fig4} shows 
that there is significant heating throughout the parameter space, but 
greater at higher ionization parameter.  Free-free heating also 
contributes.  The resulting equilibrium temperature is greater than at 
low density by approximately a factor of 10.  Another feature of the equilibrium 
temperature distribution at this density  is the disappearance of the 
thermal instability; the strong temperature-dependent cooling seen at low density is 
absent.  Accompanying this is the fact that the contours of constant heating and 
cooling are nearly congruent, in contrast to the low density behavior.  This is a
manifestation of the fact that in local thermodynamic equilibrium (LTE), the heating
and cooling balance each other identically at all temperatures, and there is 
no preferred equilibrium temperature.  At the density shown here, LTE is attained for 
the level populations and rates affecting many of the ions in the gas.  This was 
illustrated in \citet{Kall01}, Figure 10.

\subsection{Line Emissivities}

\begin{figure*}[p]
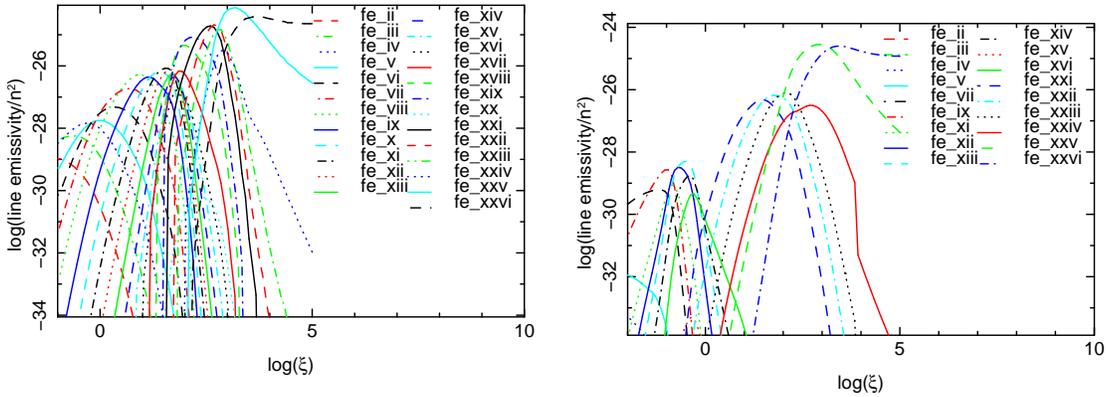
 
\includegraphics*[angle=270, scale=0.3]{fig6a.ps}
\includegraphics*[angle=270, scale=0.3]{fig6b.ps}
\caption{\label{fig6} Line emissivities vs. ionization parameter log($\xi$) at density ${\rm n}_e=10^4$ (left) and ${\rm n}_e=10^{20}$ cm$^{-3}$ (right).}
\end{figure*}

Line emissivities are affected by density in different ways according to the 
dominant process for the line emission.  Lines emitted by recombination or electron impact 
collisions depend on density according to ${\rm n}_e^2$.  Lines emitted by resonance 
scattering or fluorescence are proportional to gas density and the radiation flux, which is 
equivalent to density ${\rm n}_e^2 \xi$.  In the previous subsection we have shown that the high density  model with 
${\rm n}_e=10^{20}$ cm$^{-3}$ produces higher ionization and higher temperature than the ${\rm n}_e=10^4$ cm$^{-3}$ model.  
The generally higher ionization balance shifts the line emissivity in ionization parameter 
space.   Figure \ref{fig6} shows the emissivities in units erg cm$^3$ s$^{-1}$, i.e. emissivity 
with the density ${\rm n}_e^2$ dependence divided.  
Comparison of the panels of Figure \ref{fig6} shows that the density ${\rm n}_e^2$ scaling works in 
an approximate sense, the envelope of most of the emissivities is similar between the two 
densities.  The emissivities per ion are very similar between the two models, though the 
curves for each ion are shifted towards lower $\xi$ in the high density models.  This 
suggests that, when these rates are used to calculate the flux from a physical model for 
a X-ray illuminated slab of gas at density ${\rm n}_e=10^{20}$ cm$^{-3}$, the total line flux escaping the 
slab will be similar to a low density model, though shifted in energy owing to the higher 
degree of ionization.  In this paper, we will not present physical model slabs, since these 
depend on solutions to the radiative transfer equation, and that is beyond the scope 
of this paper.  We will present such models in a future publication.

\subsection {Opacity and Emissivity}

\begin{figure*}[p] 
\includegraphics*[angle=270, scale=0.3]{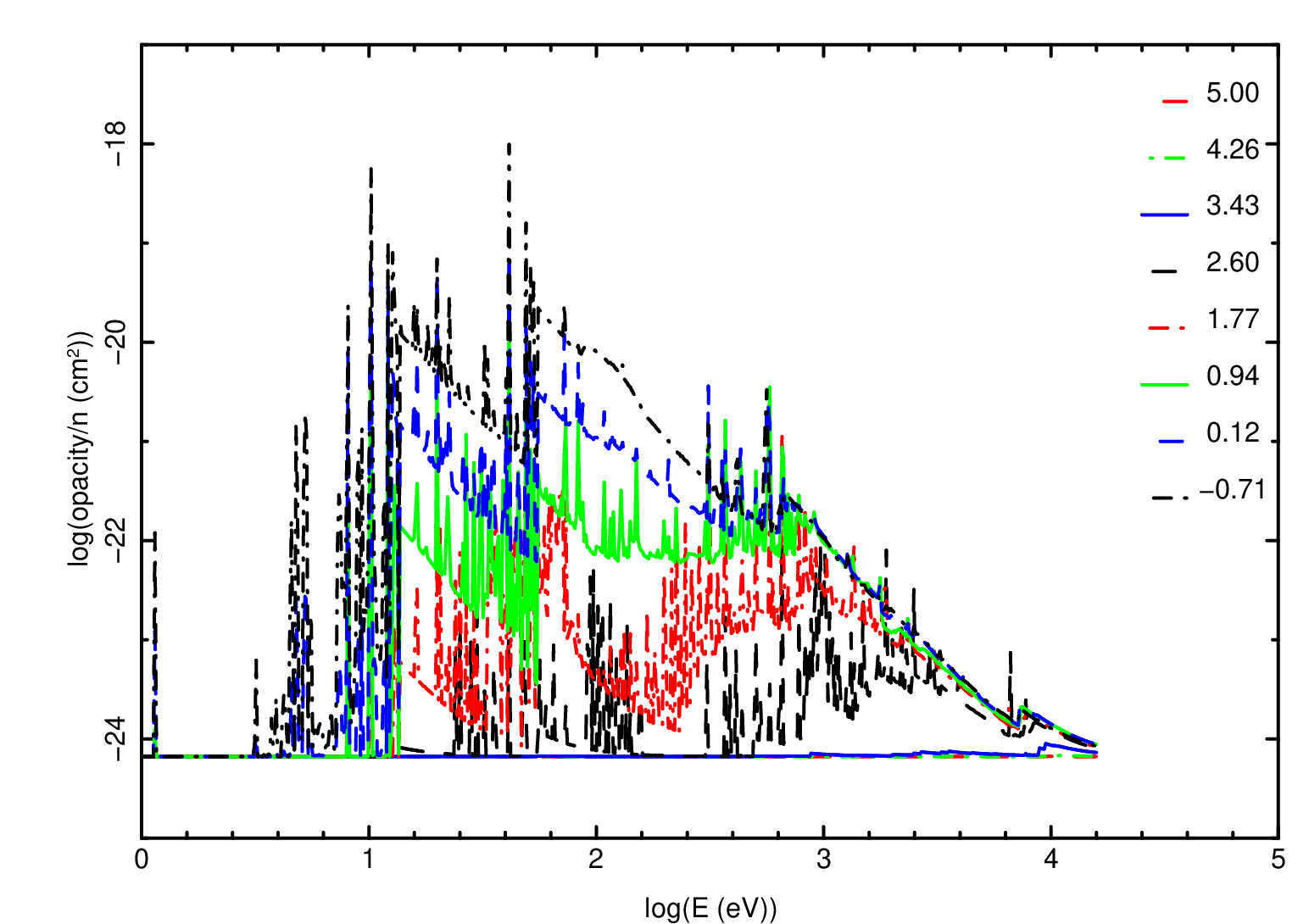}
\includegraphics*[angle=270, scale=0.3]{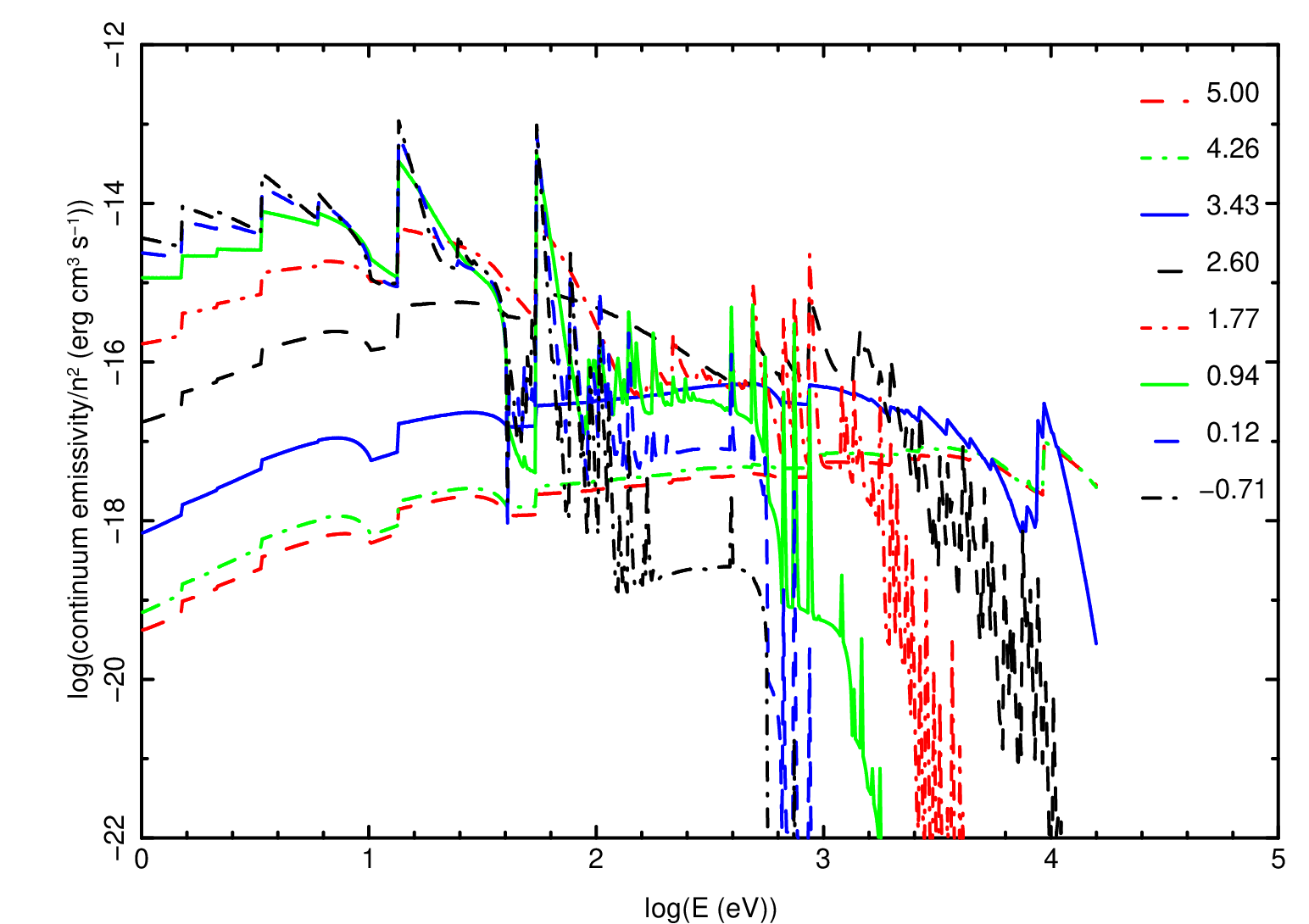}
\caption{\label{fig7} Opacity (left) and continuum emissivity (right) 
for a low density  ${\rm n}_e=10^4$ cm$^{-3}$ model for various values of ionization 
parameter shown log($\xi$) in the legend. Thomson opacity is not included.}
\end{figure*}

\begin{figure*}[p] 
\includegraphics*[angle=270, scale=0.3]{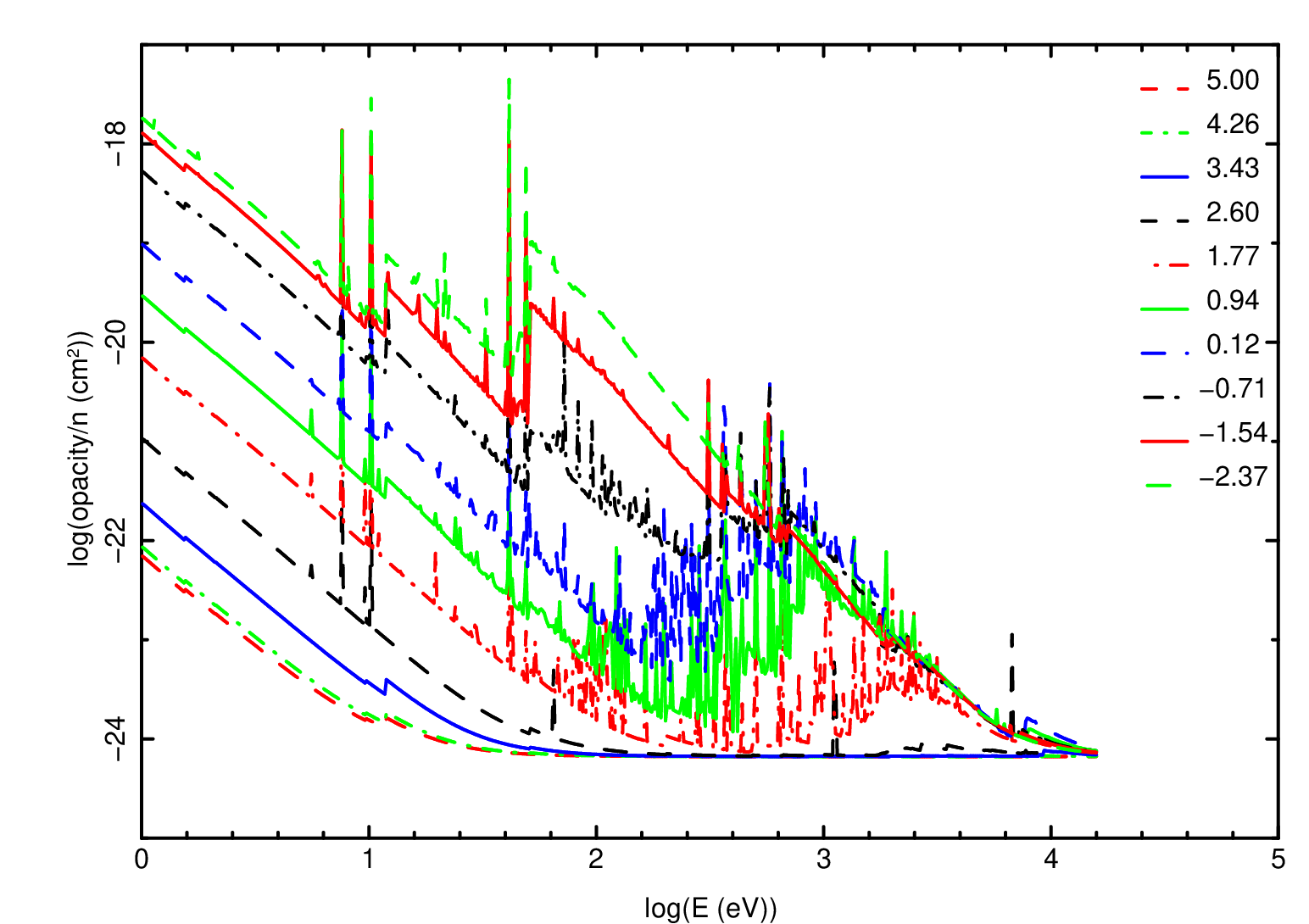}
\includegraphics*[angle=270, scale=0.3]{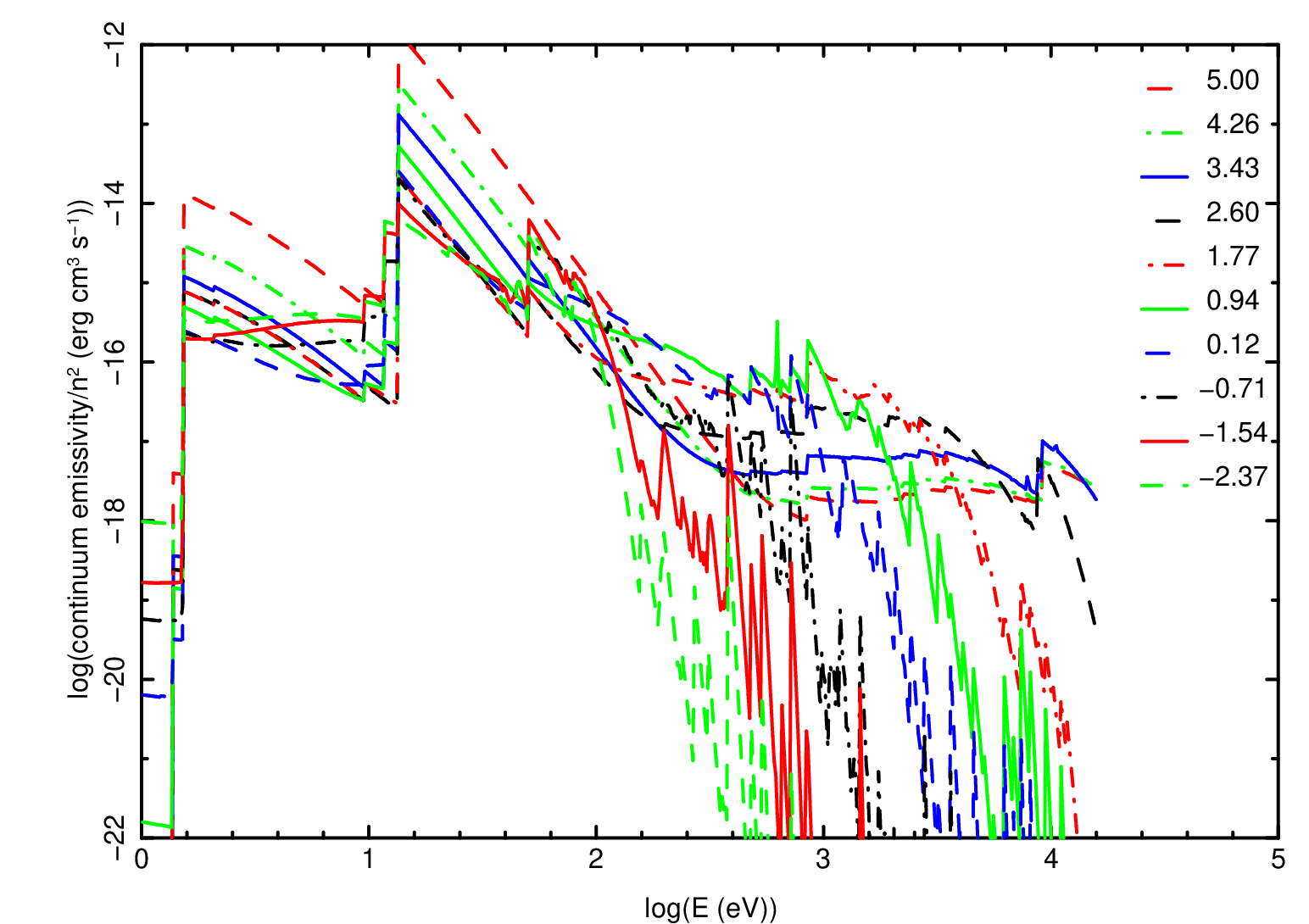}
\caption{\label{fig8} Opacity (left) and continuum emissivity (right) 
for a hith density  ${\rm n}_e=10^{20}$ cm$^{-3}$ model for various values of ionization 
parameter shown in the legend.  Thomson opacity is not included.}
\end{figure*}

The opacity of gas at high density differs from that at low density 
due to the free-free opacity.  This is apparent in comparison of Figures 
\ref{fig7} and \ref{fig8}.  The left hand panels show the opacity vs. 
energy at various ionization parameters.  The opacity is small at high 
ionization parameter, so opacity is dominated by Thomson scattering.  Photoelectric 
opacity gains in importance at lower ionization parameter, with 
significant energy dependence.  At the highest ionization parameters
edges due to highly ionized Fe and Ni are apparent.  At lower ionization 
parameter, log($\xi$)$\sim$ 0 -- 2, there is strong opacity 
between $\sim$0.5 -- 2 keV from intermediate-Z elements.  At low ionization 
parameter bound-free opacity from H and He and from valence shells of low ionization metals dominate.
At high density opacity from these ions are augmented by free-free opacity which has power 
law behavior in energy. 

Emissivity behaves in an analogous way, progressing from predominantly 
free-free at high ionization parameter to radiative recombination continua (RRCs) \citep{Lied90} at lower ionization parameter.
The low- and high density models differ at low energies, where the low 
density models shows a strong increase in total emissivity at low ionization parameters.  The 
high density models do not, owing to the fact that the ionization balance 
is changing less with ionization parameter.


\subsection {Level Populations}

At high density many of the levels of ground terms and configurations become 
populated by collisions, i.e. the collisional rates linking the levels 
become faster than the radiative decays.  This is an extension of the 
familiar process leading to population of  metastable levels in nebulae.  
Observations in many cases constrain these populations via the line emission 
or absorption indicating population of the upper or lower levels.  An 
illustration is the B-like ion Fe XXII, in which the $2s^22p$ ground 
term is split into the $j=1/2$ ground level and $j=3/2$ excited level.
The critical density leading to population of the excited levels is 
$\sim 10^{14}$ cm$^{-3}$ \citep{Mauc04}.  Above this density the line  
$2s^22p_{3/2} - 2s^23d_{5/2}$ at 11.92 $\AA$ can appear in absorption, in 
addition to the ground state line $2s^22p_{1/2} - 2s^23d_{3/2}$ at 
11.77 $\AA$.  The 11.92 $\AA$ line has been detected in astrophysical systems
including the black hole transient GROJ1655-40 \citep{Mill06}, 
and this is indicative of reprocessing in a high density medium.
It has also been used by \citet{King12} to constrain absorber density in 
in the Seyfert-1 Galaxy NGC 4051.

\begin{figure*}[p]
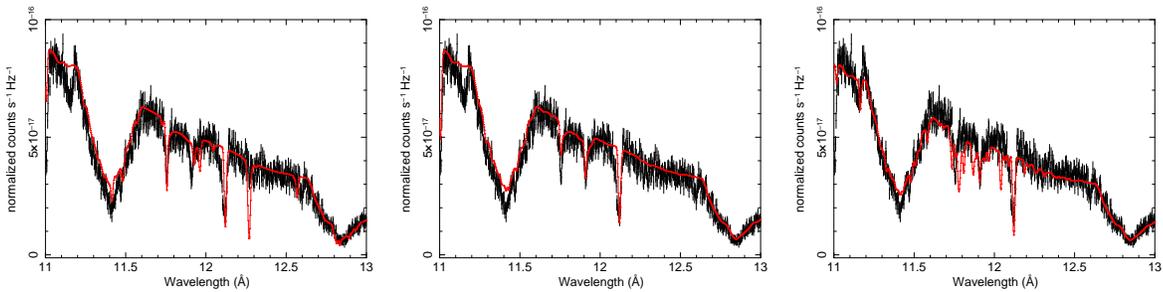
 
\includegraphics*[angle=270, scale=0.2]{fig9a.ps}
\includegraphics*[angle=270, scale=0.2]{fig9b.ps}
\includegraphics*[angle=270, scale=0.2]{fig9c.ps}
\caption{\label{fig9} Fits to the $Chandra$ HETG spectrum of GROJ1655-40 as obtained by \citet{Mill06}.  Models are at densities ${\rm n}_e = 10^4$ (left), 
$10^{16}$ (middle), and $10^{20}$ (right) cm$^{-3}$.  The effect of metastable
population on the  $2s^23d_{5/2}$ level on the 11.92$\AA$ line
 is apparent in the middle panel.
The effect of the $2s2p^2(^3D_{3/2})$ level on the 11.748 $\AA$ line 
is apparent in the right panel.}
\end{figure*}

Figure \ref{fig9} shows the spectrum predicted by {\sc xstar} in the 
region containing the Fe XXII lines, together with the $Chandra$ High 
Energy Transmission Grating (HETG) spectrum of GROJ1655-40 \citep{Mill06}.  
The three panels correspond to 
densities ${\rm n}_e = 10^4$ cm$^{-3}$, $10^{16}$ cm$^{-3}$, $10^{20}$ cm$^{-3}$.  
The data clearly shows the presence of both the 11.77 and 11.92 $\AA$ lines.
The panels show that the ${\rm n}_e = 10^4$ cm$^{-3}$ model fits the 11.77 $\AA$ line 
but fails to fit the 11.92 $\AA$ line. The ${\rm n}_e = 10^{16}$ cm$^{-3}$ model 
fits both lines qualitatively.  The ${\rm n}_e = 10^{20}$ cm$^{-3}$ model shows the 
presence of many other lines arising from excited levels which require
this density for excitation.  Notable among them is the line at 11.748 $\AA$
which arises from the $2s2p^2(^3D_{3/2})$ level.  This line is 
stronger than the 11.97 $\AA$ line at this density.  This illustrates 
that the absence of such lines in the spectrum sets an upper limit on the 
density in this source.

\section{Conclusions}
\label{discussion}

In this paper we have explored the effects of high densities
on models for astrophysical gas ionized and heated by photoionization.
Our models are valid up to density ${\rm n}_e \sim 10^{22}$ cm$^{-3}$; we have illustrated our results with 
models at ${\rm n}_e \sim 10^{20}$ cm$^{-3}$.   We have shown that:
\begin{itemize}
\item{As density is increased above ${\rm n}_e \sim 10^4$ cm$^{-3}$, the net recombination
decreases, due to the suppression of dielectronic recombination, 
leading to generally higher ionization for comparable conditions.}
\item{This trend is reversed at higher density ${\rm n}_e\geq 10^{15}$ cm$^{-3}$, due to 
the onset of three body recombination, leading to generally lower 
ionization for comparable conditions.}
\item{Lower ionization generally leads to greater photoionization heating, 
and thus higher temperature.}
\item{Stimulated recombination becomes important at high radiation intensities, 
and this accompanies high gas densities when the ionization parameter is held 
fixed.  The net effect is similar to the effect of three body recombination.}
\item{Additional heating at high density comes from free-free heating, again 
due to the greater local radiation intensity.}
\item{When the effective ionization potential is reduced due to Debye screening 
to the point where many excited levels are in the continuum, the behavior 
of many ions changes qualitatively.  The net recombination rate is again 
reduced, and line emission and cooling efficiencies are also reduced.}
\item{Many of these processes depend on the effective charge of the ion: higher
densities are needed to make three body recombination important for highly 
charged ions compared with near-neutrals; continuum lowering effects are 
more important for low charge ions, compared with highly charged ions at 
a given density.}
\item{Comparison of ionization and thermal balance between low (${\rm n}_e=10^{4}$ cm$^{-3}$) and high density
(${\rm n}_e=10^{20}$ cm$^{-3}$) photoionization models shows that the latter are hotter 
by up to a factor of 10, and significantly more highly ionized for a given 
ionization parameter.  Thermal instability at constant pressure does not occur.}
\item{Line emissivities and opacities generally obey simple scaling behavior, 
but there are important departures which will affect model spectra.}
\item{Densities greater than those considered previously lead to excitation 
of new metastable levels and accompanying line formation.}
\end{itemize}

The {\sc xstar} atomic rate coefficients and code calculating ionization 
balance, atomic level populations, temperature, emissivity and opacity
is also used in calculation of the reflection spectra from 
Compton-thick atmospheres  with the {\sc xillver} code by \cite{Garc10,Garc11,Garc13}.  
These models also allow treatment of angle dependence of the 
radiation field, both the effect of non-normal incident radiation 
and also the angle dependence of the reflected radiation.
In subsequent papers we will present models 
at densities appropriate to astrophysical sources which exhibit reflection 
spectra.  These will include all the ingredients of the \cite{Garc13} models,
and span similar free parameter values,  plus include 
the revisions to atomic rate coefficients described so far.


\appendix

\section{Appendix}

It is worthwhile to compare the particular effects of the 
density, and other recent changes to predictions of {\sc xstar}
to typical variations between the predictions of other 
models and codes designed for solving similar problems. 
These have been described in the proceedings of the series 
of non-LTE workshops, and in the publications by \citet{ralc16,hans20}.  Most
of these codes have been applied extensively to terrestrial plasmas, such 
as laser-produced plasmas.  Comparisons with those codes necessarily 
focusses on elements which are of interest to astrophysics.  
We have compared {\sc xstar} results with two of the models described in 
the compilation of model results by \citet{hans20}.  

\begin{figure*}[p] 
\includegraphics*[angle=0, scale=1.]{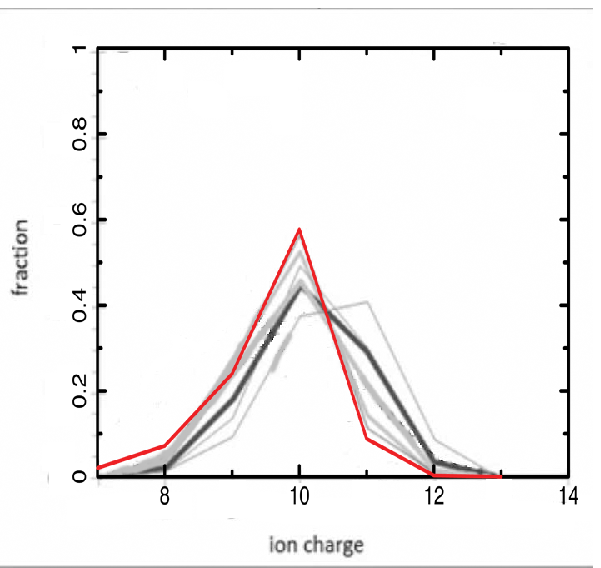}
\includegraphics*[angle=0, scale=1.]{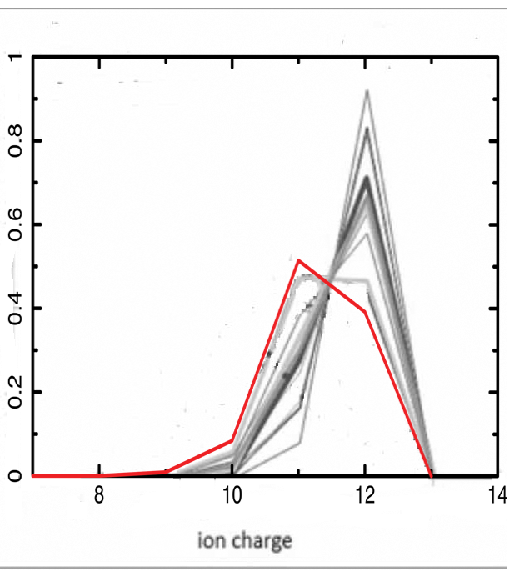}
\caption{\label{figa1} Comparison of charge state distribution between the ensemble of models described in \cite{hans20} for silicon with photoionization (grey curves) and an {\sc xstar} model for similar conditions (red curve).  Right panel:  undiluted blackbody illumination; Left panel: diluted blackbody illumination by 10x}
\end{figure*}

The first comparison is a model for Si produced primarily by photoionization,
denoted 'steady state Si' in that paper.  The ionizing spectrum is chosen to 
be a blackbody with kT=63 eV, with flux specified as either the full blackbody 
or else diluted by 10x.  We compare with the models with density $10^{19}$ cm$^{-3}$, which then corresponds to ionization parameter values log($\xi$)=1.3 or 0.3.  The resulting ionization balance of these models are shown in figure 
\ref{figa1}.  The ensemble of other code results are shown as grey curves and 
{\sc xstar} is shown in red.  This shows that our models produce an 
ionization balance which is slightly less than most other models for the 
high-$\xi$ model but is very similar to the other  models for the 
low-$\xi$ model.

\begin{figure*}[p] 
\includegraphics*[angle=0, scale=1.]{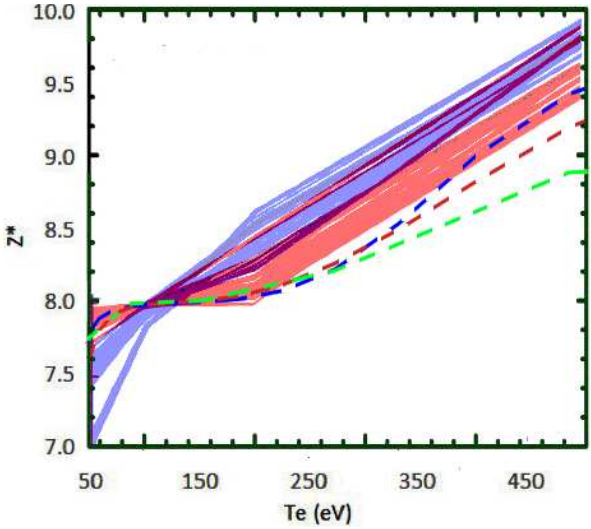}
\caption{\label{figa2} Comparison of mean ion charge vs. temperature between the ensemble of models described in \cite{hans20} for neon with coronal conditions.(solid curves) and {\sc xstar} models for similar conditions (dashed curves). Colors 
correspond to density  10$^{19}$ cm$^{-3}$ (blue) 10$^{21}$ cm$^{-3}$ (red). 
Density  10$^{4}$ cm$^{-3}$ (green) is shown for {\sc xstar} only.}
\end{figure*}

The second comparison is a model for Ne produced by collisional ionization, 
denoted 'steady state Ne' by \citet{hans20}.  We compare with the models with density $10^{19}$ cm$^{-3}$ and  $10^{21}$ cm$^{-3}$. The resulting mean charge
vs. temperature is shown in figure \ref{figa2}.  
The ensemble of other code results are shown as grey curves and 
{\sc xstar} is shown in red.  This shows that our models produce 
mean ion charge which is less than the average of the 
other models by 0.2 dex for both densities, though we note the dispersion
among the other results is larger than this.  We also note that the change 
in the mean charge between the two densities is also $\sim$0.2 dex, and that 
this is similar to the behavior of the mean of the models shown in 
\citet{hans20}.  The  {\sc xstar} results are derived from the rates from 
\citet{brya06} at low density, and agree with those results identically 
at density 10$^4$ cm$^{-3}$.  At higher density the {\sc xstar} results predict
lower mean ion charge than the models cited in the \citet{hans20} 
compilation by about 0.1 dex on average.  One possible reason for this is the
treatment of electron impact excitation from metastable levels 
to doubly excited levels, which then autoionize.   This process is included
in the low density rates from \citet{brya06} for collisions from the ground 
state, but {\sc xstar} likely has fewer such transitions included for
metastables states which are populated at density $\geq 10^{19}$ cm$^{-3}$.

\acknowledgements

Partial support for this work was provided by grant 80NSSC17K0345 through the NASA APRA program. J.A.G. acknowledges support from NASA ADAP grant 80NSSC19K0586, and from the Alexander von Humboldt Foundation. JD is Research Fellow of the Belgian Fund for Research Training in Industry and Agriculture (FRIA) while PP and PQ are, respectively, Research Associate and Research Director of the Belgian Fund for Scientific Research (F.R.S.-FNRS).
AO is supported by NASA under award number 80GSFC17M0002.  We are very grateful
to the referee for several constructive  suggestions.

%
%
%
%
%
%

\end{document}